\documentclass[10pt,preprint]{aastex}
\usepackage{graphicx,apjfonts,emulateapj5,onecolfloat5}
\usepackage{subfigure}
\usepackage{ulem}
\usepackage{epsf}
\usepackage{dcolumn}   % Align table columns on decimal point
\usepackage{bm}        % bold math
\usepackage{multirow}

\def\apj{Astrophys. J.~}
\def\apjl{Astrophys. J. Lett.~}

\def\apjs{Astrophys. J. Suppl. Ser.~}

\def\jcap{J.~Cosmol.~Astropart.~Phys.~}

\def\prd{Phys. Rev. D~}

\shorttitle{Constraints on non-flat cosmologies}
\shortauthors{Chen, Y. et al.}

\begin{document}
\twocolumn[
%\onecolumn[

%% LaTeX will automatically break titles if they run longer than
%% one line. However, you may use \\ to force a line break if
%% you desire.

\title{Constraints on non-flat cosmologies with massive neutrinos after Planck 2015}

%% Use \author, \affil, and the \and command to format
%% author and affiliation information.
%% Note that \email has replaced the old \authoremail command
%% from AASTeX v4.0. You can use \email to mark an email address
%% anywhere in the paper, not just in the front matter.
%% As in the title, you can use \\ to force line breaks.

\author{Yun Chen\altaffilmark{1},
        Bharat Ratra \altaffilmark{2},
        Marek Biesiada \altaffilmark{3,4},
        Song Li \altaffilmark{5},
        and Zong-Hong Zhu \altaffilmark{3}
        }

%% Mark off your abstract in the ``abstract'' environment. In the manuscript
%% style, abstract will output a Received/Accepted line after the
%% title and affiliation information. No date will appear since the author
%% does not have this information. The dates will be filled in by the
%% editorial office after submission.

\begin{abstract}
%%%%%%%%%%%%%%%%%%%%%%%%%%%%%%%%%%%%%%%%%%%%%%%%%%%%%%%%%%%%%%%%%%%
We investigate two dark energy cosmological models (i.e., the $\Lambda$CDM
and $\phi$CDM models) with massive neutrinos assuming two different neutrino mass hierarchies in both the spatially flat and
non-flat scenarios, where in the $\phi$CDM model the scalar field possesses
an inverse power-law potential, $V(\phi)\propto {\phi}^{-\alpha}$ ($\alpha>0$).
Cosmic microwave background data from Planck 2015, baryon acoustic oscillations
data from 6dFGS, SDSS-MGS, BOSS-LOWZ and BOSS CMASS-DR11, the JLA compilation
of Type Ia supernova apparent magnitude observations, and the Hubble Space
Telescope $H_0$ prior, are jointly employed to constrain the model parameters. We first determine constraints assuming three species of degenerate massive neutrinos.
In the spatially flat (non-flat) $\Lambda$CDM model, the sum of neutrino
masses is bounded as $\Sigma m_{\nu} < 0.165 (0.299)$ eV at 95\% confidence level (CL).
Correspondingly, in the flat (non-flat) $\phi$CDM model, we find
$\Sigma m_{\nu} < 0.164 (0.301)$ eV at 95\% CL. The
inclusion of spatial curvature as a free parameter results in a significant
broadening of confidence regions for $\Sigma m_{\nu}$ and other
parameters. In the scenario where the total neutrino mass is dominated by the heaviest neutrino mass eigenstate, we can obtain the
similar conclusions as those obtained in the degenerate neutrino mass scenario. In addition,
the results show that the bounds on $\Sigma m_{\nu}$ based on two
 different neutrino mass hierarchies have insignificant differences
in the spatially flat case for both the $\Lambda$CDM and $\phi$CDM models, however, the corresponding differences are
larger in the non-flat case.

\end{abstract}

%% Keywords should appear after the \end{abstract} command. The uncommented
%% example has been keyed in ApJ style. See the instructions to authors
%% for the journal to which you are submitting your paper to determine
%% what keyword punctuation is appropriate.

]

\altaffiltext{1}{Key Laboratory for Computational Astrophysics, National Astronomical Observatories, Chinese Academy of Sciences, Beijing, 100012, China; chenyun@bao.ac.cn}

\altaffiltext{2}{Department of Physics, Kansas State University, 116 Cardwell Hall, Manhattan, KS 66506, USA}
\altaffiltext{3}{Department of Astronomy, Beijing Normal University, Beijing 100875, China}
\altaffiltext{4}{Department of Astrophysics and Cosmology, Institute of Physics, University of Silesia, Uniwersytecka 4, 40-007 Katowice, Poland}
\altaffiltext{5}{Department of Physics, Capital Normal University, Beijing 100048, China}

%% From the front matter, we move on to the body of the paper.
%% In the first two sections, notice the use of the natbib \citep
%% and \citet commands to identify citations.  The citations are
%% tied to the reference list via symbolic KEYs. The KEY corresponds
%% to the KEY in the \bibitem in the reference list below. We have
%% chosen the first three characters of the first author's name plus
%% the last two numeral of the year of publication as our KEY for
%% each reference.

\section{INTRODUCTION}
%%%%%%%%%%%%%%%%%%%%%%%%%%%%%%%%%%%%%%%%%%%%%%%%%%%%%%%%%%%%%%%%%%%
\label{intro}

To date, there is firm evidence for neutrino oscillations (see the reviews:
Maltoni et al. 2004; Fogli et al. 2006; Balantekin \& Haxton 2013) from
measurements on solar (Ahmad et al. 2001), atmospheric (Fukuda et al. 1998),
reactor (An et al. 2012; Ahn et al. 2012) and accelerator beam (Agafonova et
al. 2010) neutrinos. These measurements imply that neutrinos have small but
non-zero masses, with at least two species being non-relativistic today.
Experiments have placed restrictive limits on differences
of two squared neutrino masses, such as $\Delta m_{21}^2=m_2^2-m_1^2
\sim 8\times 10^{-5}$eV$^2$ (Abe et al. 2008) and $\Delta m_{32}^2
=m_3^2-m_2^2\sim 3\times 10^{-3}$eV$^2$ (Ashie et al. 2005), but give no
constraint on their absolute mass scales.  Here $m_1$, $m_2$ and $m_3$ denote the
masses of neutrino mass eigenstates. The measurement of the absolute neutrino mass scale remains a big challenge for both experimental particle physics and observational cosmology. Fortunately, a variety of cosmological probes can provide the crucial complementary information on absolute neutrino mass scale. Current cosmological data can provide an upper limit on the total neutrino mass $\sum m_{\nu}$ (summed over the three neutrino families) of order 1 eV or less (Lesgourgues \& Pastor 2012), though they are not very sensitive to the neutrino mass hierarchy.

Massive neutrinos are the only particles that have undergone the transition from radiation to
matter as the universe expanded and cooled (Lesgourgues \& Pastor 2006). Before the
non-relativistic transition the neutrinos behave like radiation. Thus, when
the total neutrino mass $\Sigma m_{\nu}$ increases, there is more relativistic matter at early times and the matter-radiation equality occurs later, so the scale factor at the epoch of matter-radiation equality  $a_{eq}$ increases (i.e., $z_{eq}$ gets lower). The cosmic microwave background (CMB)
radiation and large-scale structure (LSS) distributions are very sensitive
to $a_{eq}$, which provides potential ways to constrain $\Sigma m_{\nu}$
through CMB and LSS observations. In addition, the massive neutrinos are
non-relativistic today, so they contribute to the recent expansion rate of
the universe like cold dark matter. Moreover, after thermal decoupling the
massive neutrinos freely stream a distance called the free-streaming length.
This disrupts the structure formation on scales below the free-streaming
length. Because of the above effects, massive neutrinos can leave imprints
on cosmological observables. This is why a variety of cosmological tests are
sensitive to the absolute scale of neutrino mass, such as the CMB anisotropy,
galaxy, and Lyman-alpha forest distributions as well as the distance
information from baryon acoustic oscillations (BAO) and type Ia supernovae
(SNe Ia) measurements.

The limits on $\sum m_{\nu}$ obtained from cosmology, so far, are rather model
dependent and vary strongly with the data combination adopted. In Hannestad (2005), it was found that when the dark energy equation of state (EoS) is taken as a free (but constant) parameter, the cosmological bound on $\sum m_{\nu}$
 is relaxed by more than a factor of two, to $\sum m_{\nu} < 1.48$ eV (95\% CL), compared with $\sum m_{\nu} < 0.65$ eV (95\% CL) in the $\Lambda$CDM model. The above results were obtained from a combination of CMB measurements from the first-year Wilkinson Microwave Anisotropy Probe (WMAP) observations (Bennett et al. 2003), the galaxy power spectrum based on the Sloan Digital Sky Survey (SDSS) Data Release 2 (Tegmark et al. 2004), the SNe Ia data from Riess et al. (2004), and the $H_0$ prior
 from the Hubble Space Telescope (HST) Key Project with $H_0 = 72\pm8$ km s$^{-1}$ Mpc$^{-1}$ (Freedman et al. 2001).  The two models studied in Hannestad (2005) were also constrained in Wang et al. (2012) with updated cosmological data, where the corresponding results turned out to be $\sum m_{\nu} < 0.627$ (95\% CL) for an arbitrary (but constant) EoS and $ \sum m_{\nu} < 0.476$ eV (95\% CL) for the $\Lambda$CDM model.
Based on the benefits of the more precise cosmological data, the bound on $\sum m_{\nu}$ is much more restrictive for each individual model,
 and the difference of the bounds on $\sum m_{\nu}$ from the two models is also reduced. The bound on $\sum m_{\nu}$ in the framework of time evolving EoS, $\omega(z) = \omega_0 + \omega_1*z/(1+z)$, was also investigated in the literature (Xia et al. 2007; Xia et al. 2008; Li et al. 2012), and revealed the degeneracy between $\sum m_{\nu}$ and the EoS $\omega$ parameters. In Smith et al. (2012), it was found that with non-vanishing curvature density parameter $\Omega_k \neq 0$ the 95\% upper limit on $\sum m_{\nu}$ was more than double with respect to the case of a flat universe.  This implies the strong degeneracy between curvature and $\sum m_{\nu}$.

In this paper, we present constraints on the total mass of ordinary (active) neutrinos $\sum m_{\nu}$ assuming no extra relics.
Current cosmological data are not yet sensitive to the mass of individual neutrino species, i.e. the mass hierarchy.
Under this situation, two scenarios for the mass splitting of the standard three flavor neutrinos are often used in cosmology:
 (i) assuming three species of degenerate massive neutrinos, neglecting the small differences
in mass expected from the observed mass splittings; and (ii) assuming the total neutrino mass dominated by the heaviest neutrino
 mass eigenstate (i.e. two massless and one massive neutrino).
We will analyze and compare the constraints based on both
the $\Lambda$CDM and $\phi$CDM models in both the spatially flat ($\Omega_k=0$)
and non-flat ($\Omega_k\neq0$) cases taking into account two different mass hierarchies.
The $\phi$CDM model --- in which dark energy is modeled as a scalar
field $\phi$ with a gradually decreasing (as a function of $\phi$) potential $V(\phi)$ --- is a simple dynamical model with
dark energy density slowly decreasing in time. This model could resolve some of the puzzles of
the $\Lambda$CDM model, such as the coincidence and fine-tuning problems
(Peebles \& Ratra 1988; Ratra \& Peebles 1988). Here we focus on an inverse power-law potential $V(\phi)
\propto \phi^{-\alpha}$, where $\alpha$ is a nonnegative constant
(Peebles \& Ratra 1988; Ratra \& Peebles 1988). When $\alpha = 0$ the
$\phi$CDM model is reduced to the corresponding $\Lambda$CDM scenario. The $\phi$CDM model with this kind of $V(\phi)$
has been extensively investigated, mostly in the spatially flat case
(Chen et al. 2015; Avsajanishvili et al. 2014, 2015; Lima et al. 2015;
Pavlov et al. 2014; Farooq et al. 2013a, 2013b; Farooq \& Ratra 2013;
Chen \& Ratra 2011;  Samushia \& Ratra 2010; Samushia et al 2007;
Chae et al. 2004; Chen \& Ratra 2004; Podariu \& Ratra 2000), and only a
limited attention has been paid to the non-flat scenario (Pavlov et al. 2013;
Farooq et al. 2015; Gosenca \& Coles 2015). However, the above mentioned
literature on the $\phi$CDM model did not consider massive neutrinos. In our previous work the $\phi$CDM model with massive neutrinos has been studied under the assumption of spatial flatness (Chen \& Xu 2016) using a combination of CMB data from Planck 2013 and other datasets. In this work, the $\phi$CDM model with massive neutrinos will be further investigated in both flat and non-flat scenarios by using a combination of the CMB data from Planck 2015,
 BAO data from 6dFGS, SDSS-MGS, BOSS-LOWZ and CMASS-DR11, the JLA compilation of SNe Ia observations, and two different $H_0$ priors.

The rest of the paper is organized as follows. Constraints from the cosmological data are derived in Sec.\
{\ref{Observation}}, and the results for $\phi$CDM model are compared with those for the $\Lambda$CDM model in both the spatially flat and non-flat scenarios. We summarize our
 main conclusions in Sec.\ {\ref{summary}}.

%%%%%%%%%%%%%%%%%%%%%%%%%%%%%%%%%%%%%%%%%%%%%%%%%%%%%%%%%%%%%%%%%%%%%%%%%%%%%%%%%%%%%%%%%%%%%%%%%%%%%%%%%%%%%%%%%%%%%%%%%%%%%%%%

\section{Observational constraints}
\label{Observation}

We consider four cosmological models with massive neutrinos in this paper, i.e., (i) the spatially flat $\Lambda$CDM model,
(ii) the spatially non-flat $\Lambda$CDM model, (iii) the spatially flat $\phi$CDM model, and (iv) the spatially non-flat $\phi$CDM model.
And for each of the four models, we take into account two different scenarios for the neutrino mass hierarchy as mentioned above.
Evolution of the background and perturbations are both considered within the linear perturbation theory.  Appropriate formulae for the
$\Lambda$CDM and $\phi$CDM models in the spatially flat scenario are presented in Section 2 of  Chen \& Xu (2016). It is easy to generalize
them to the non-flat scenario by inclusion the curvature term $\Omega_k$. The parameter spaces of the models under consideration are as follows:

\begin{equation}
\label{eq:P_flatlcdm}
\textbf{P}_{1} \equiv
 \{\Omega_b h^2, \Omega_c h^2, 100\theta_{MC}, \tau, {\rm{ln}}(10^{10} A_s), n_s, \Sigma m_\nu\},
\end{equation}

\begin{equation}
\label{eq:P_nonflatlcdm}
\textbf{P}_{2} \equiv
 \{\Omega_b h^2, \Omega_c h^2, 100\theta_{MC}, \tau, {\rm{ln}}(10^{10} A_s), n_s, \Sigma m_\nu, \Omega_k\},
\end{equation}

\begin{equation}
\label{eq:P_flatphicdm}
\textbf{P}_{3} \equiv \{\Omega_b h^2, \Omega_c h^2, 100\theta_{MC}, \tau, {\rm{ln}}(10^{10} A_s), n_s, \Sigma m_\nu, \alpha\},
\end{equation}

\begin{equation}
\label{eq:P_nonflatphicdm}
\textbf{P}_{4} \equiv \{\Omega_b h^2, \Omega_c h^2, 100\theta_{MC}, \tau, {\rm{ln}}(10^{10} A_s), n_s, \Sigma m_\nu, \alpha, \Omega_k\} ,
\end{equation}
where $\textbf{P}_{1}$ and $\textbf{P}_{2}$ are the parameter spaces of $\Lambda$CDM model in the spatially flat and non-flat scenarios, respectively; $\textbf{P}_{3}$ and $\textbf{P}_{4}$ are the corresponding ones for $\phi$CDM model in the flat and non-flat scenarios. Present day densities of the baryon and cold dark matter are denoted by $\Omega_b h^2$ and $\Omega_c h^2$, respectively, $\theta_{MC}$ is an approximation to the angular size
of the sound horizon at the time of decoupling $\theta_* = r_s(z_*)/D_A(z_*)$ built in the CosmoMC package
which is based on fitting
formulae given in  Hu \& Sugiyama (1996), $\tau$ refers to the Thomson scattering optical depth due to reionization,
 $n_s$ and $A_s$ are the power-law index and amplitude of the power-law scalar primordial power spectrum of scalar perturbations,
$\Sigma m_\nu$ is the sum of neutrino masses, $\Omega_k$ is the dimensionless spatial curvature density today, and $\alpha$ determines the steepness of the scalar field potential in the framework of $\phi$CDM model.

\subsection{Cosmological data sets}

According to the constraints from the current cosmological observations the value of $\Sigma m_{\nu} \lesssim 1$ eV.
This is below the limit to which the CMB power spectrum (excluding the late-time gravitational lensing effect on the power spectrum) alone
 can be sensitive (Komatsu et al. 2009). In other words, the massive neutrinos are relativistic at the decoupling epoch, so the effect of the
  massive neutrinos in the primary CMB  power spectrum is very small. The main effect is around the first acoustic peak and is due to the early
  integrated Sachs-Wolfe (ISW) effect. After the relativistic to non-relativistic transition, the massive
neutrinos behave like cold matter. However, the non-relativistic massive neutrinos can suppress the CMB lensing potential on scales smaller than the horizon size. Thus CMB lensing is a useful probe for massive neutrinos. The CMB dataset adopted here is a combination of the low multipoles ($l=2-29$) joint TT, EE, BB and TE likelihood, and high multipoles joint TT ($l =30-2508$), TE ($l=30-1996$), and EE ($l=30-1996$) likelihood, along with CMB lensing ($l=40-400$) likelihood from Planck 2015 (Adam et al. 2015; Ade et al. 2015). BAO data from galaxy redshift surveys are a powerful cosmological probe, that can supply the Hubble expansion rate and angular diameter distance at different redshifts. The BAO dataset employed here is a combination of measurements from the 6dFGS at $z_{\textrm{eff}} = 0.1$ (Beutlerf et al. 2011), the SDSS Main Galaxy Sample (MGS) at $z_{\textrm{eff}} = 0.15$ (Ross et al. 2014), the Baryon Oscillation Spectroscopic Survey (BOSS) `LOWZ' sample at $z_{\textrm{eff}} = 0.32$ and BOSS CMASS-DR11 anisotropic BAO measurements at $z_{\textrm{eff}} = 0.57$ (Anderson et al. 2014). Another important cosmological probe is offered by SNe Ia, which provided the first direct evidence for cosmic acceleration. The SNe Ia sample used here is the  ``joint light-curve analysis'' (JLA) compilation of SNe Ia (Betoule et al. 2014), which is a joint analysis of SNe Ia observations including several low-redshift samples ($z < 0.1$), all three seasons from the SDSS-II ($0.05 < z < 0.4$), three years from SNLS ($0.2 < z < 1$), and 14 very high redshift ($0.7 < z < 1.4$) from the HST observations. It totals 740 spectroscopically confirmed SNe Ia with high quality light curves.
The Riess et al. (2011) HST Cepheid + SNe Ia  based estimate of $H_0 = (73.8\pm
2.4)$ km s$^{-1}$ Mpc$^{-1}$ is also used as a supplementary ``$H_0$-prior''.
Another prior is the median statistics estimate of $H_0 = (68\pm
2.8)$ km s$^{-1}$ Mpc$^{-1}$ of Chen \& Ratra (2011), which is more consistent with $H_0$ values
estimated using CMB and BAO data (e.g., Sievers et al. 2013; Aubourg et al.
2015; also see Calabrese et al. 2012).

\subsection{Results and analysis}
\label{Results}
In our analysis, the likelihood is assumed to be Gaussian, thus we have the total likelihood
\begin{equation}
\label{eq:LH_total}
\mathcal{L} \propto e^{-\chi_{\textrm{tot}}^2/2},
\end{equation}
where $\chi_{\textrm{tot}}^2$ is constructed as
\begin{equation}
\label{eq:chi2_total}
\chi_{\textrm{tot}}^2 = \chi^2_{\textrm{CMB}} + \chi^2_{\textrm{BAO}} + \chi^2_{\textrm{SNe}} + \chi^2_{H_0},
\end{equation}
with $\chi^2_{\textrm{CMB}}$, $\chi^2_{\textrm{BAO}}$, $\chi^2_{\textrm{SNe}}$ and $\chi^2_{H_0}$ denoting the contributions from CMB,
 BAO, SNe Ia and HST or median statistics $H_0$ prior data sets described
above, respectively. We derive the posterior probability distributions of parameters with Markov Chain Monte Carlo (MCMC) exploration
using the July 2015 version of CosmoMC (Lewis \& Bridle 2002).

\begin{figure*}[t]
\centering $\begin{array}{ccc}
\includegraphics[width=0.32\textwidth]{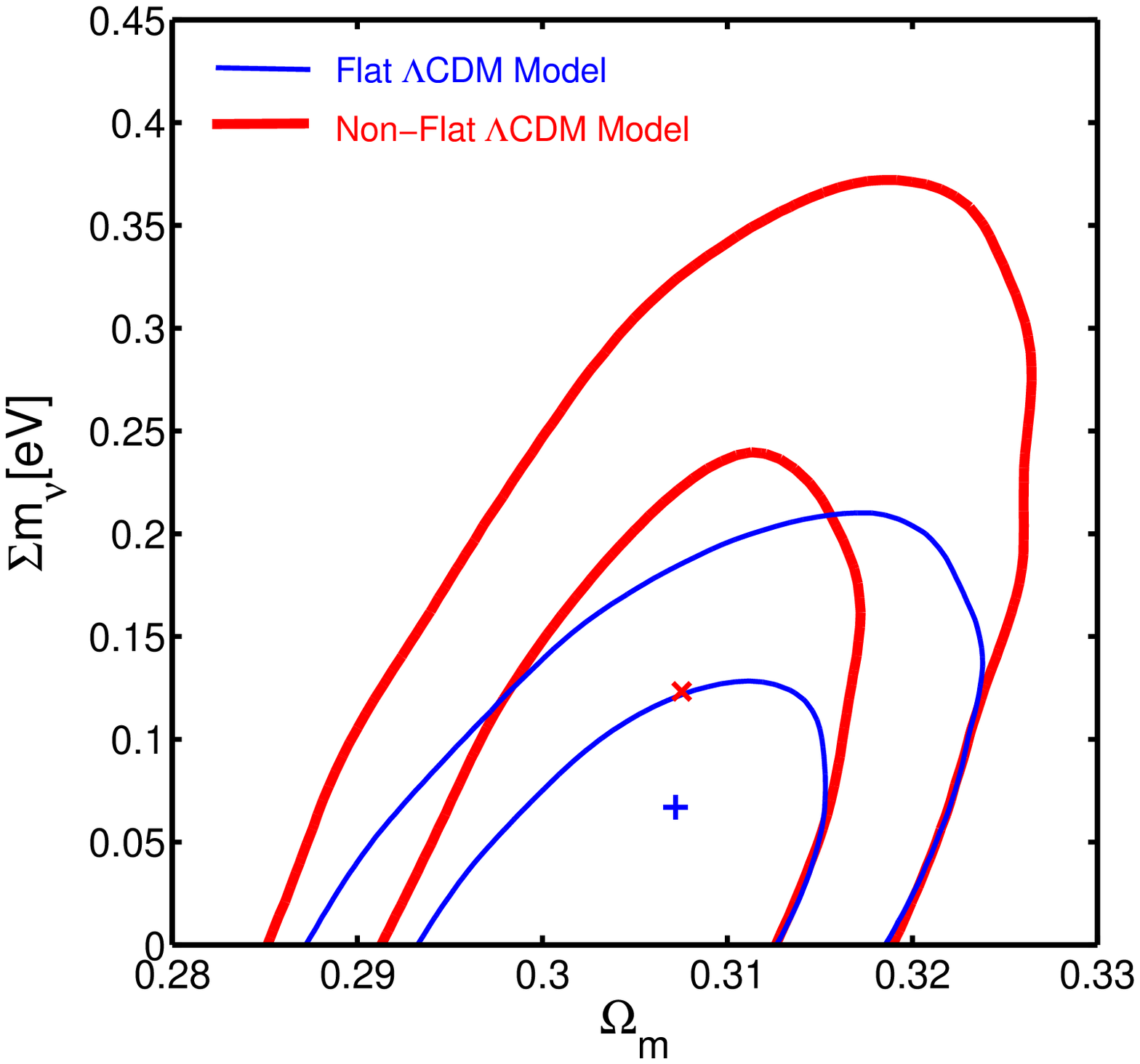}
\includegraphics[width=0.32\textwidth]{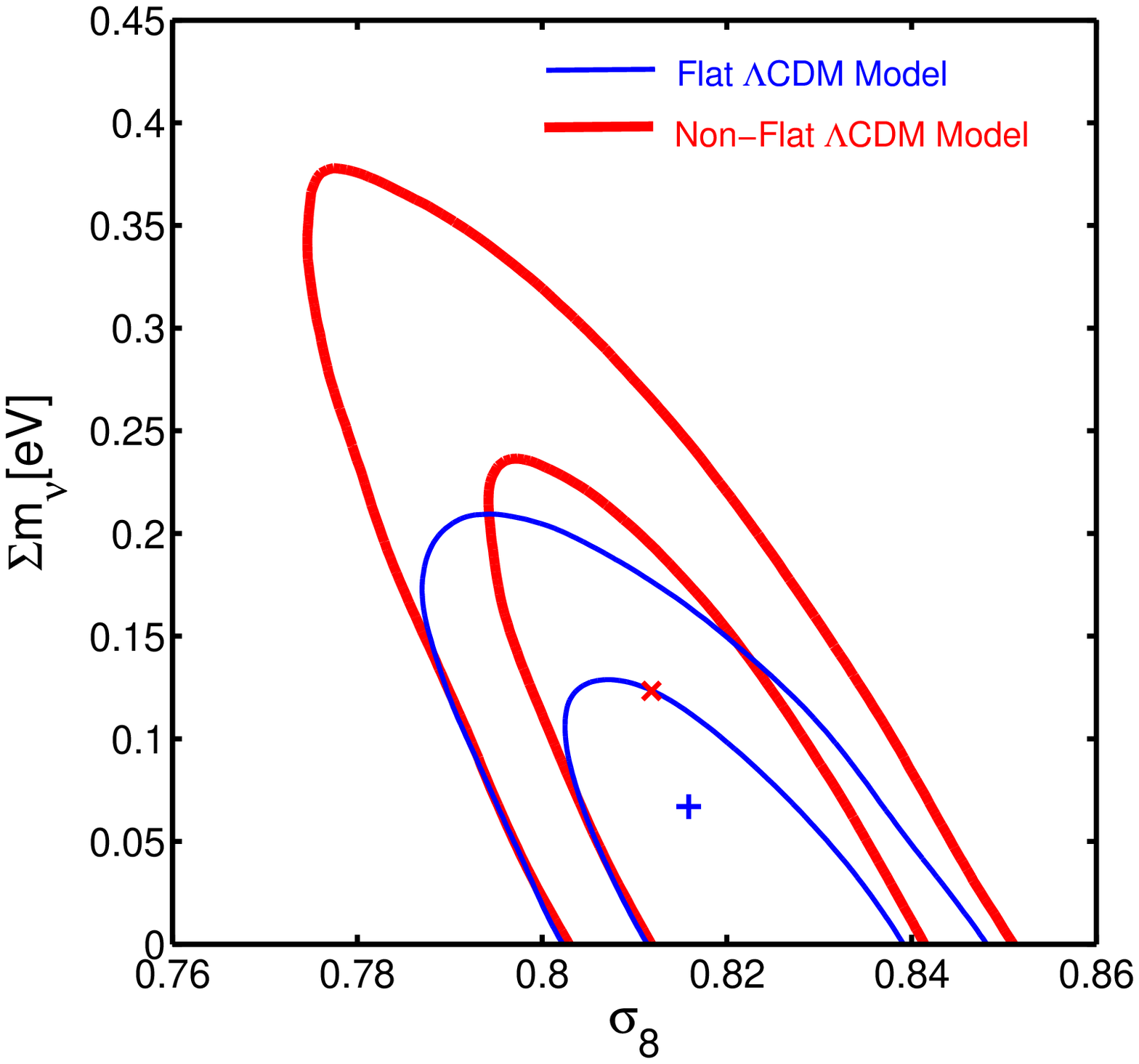}
\includegraphics[width=0.32\textwidth]{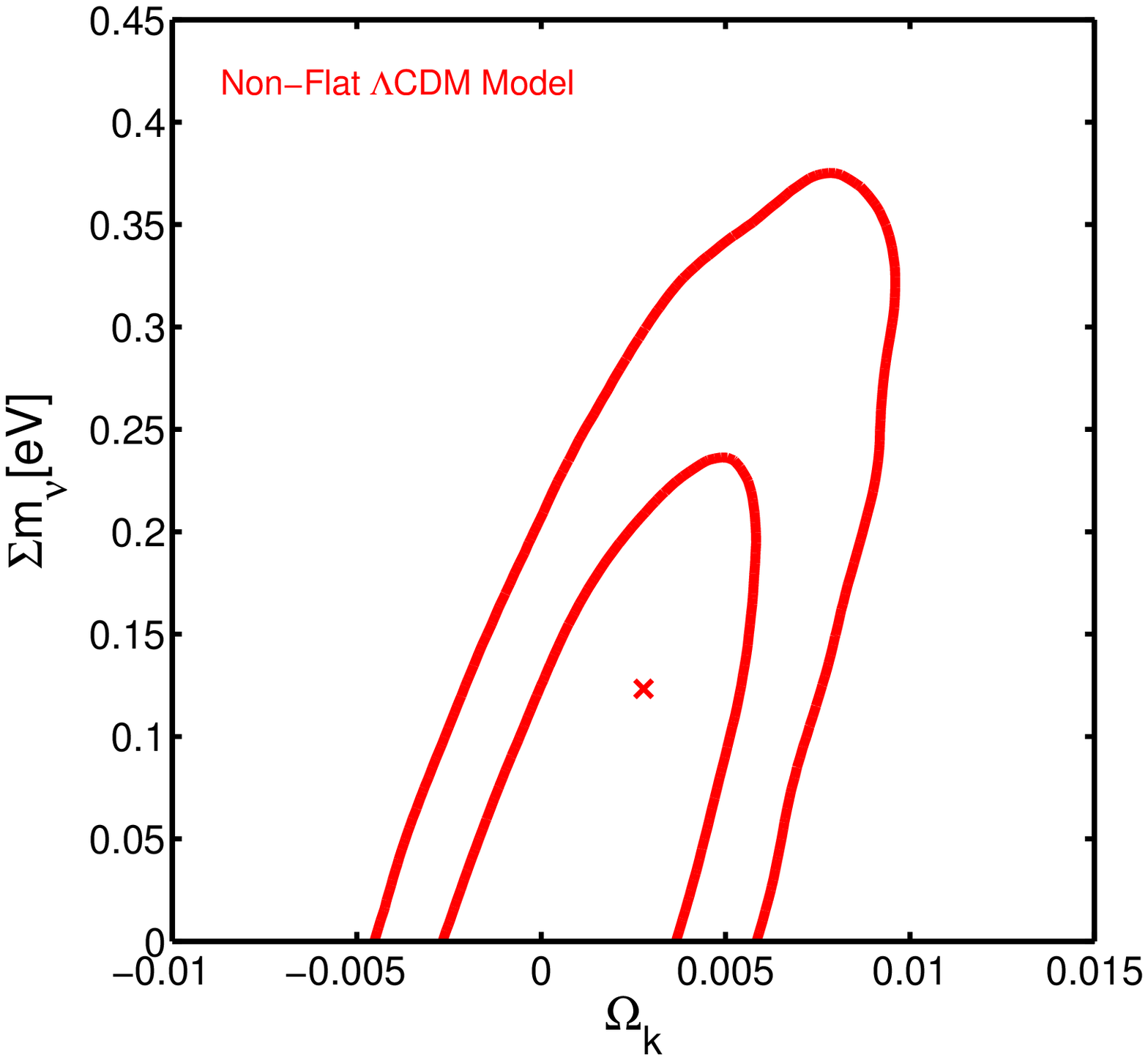}
\end{array}$
 \caption{
Contours refer to the marginalized likelihoods at 68\% and 95\% confidence levels constrained from the
joint analysis using the HST $H_0$ prior for the $\Lambda$CDM model
in the scenario assuming three species of degenerate massive neutrinos.
 \emph{\textbf{Left and middle panels}}:
contours in the $(\Omega_m, \Sigma m_{\nu})$ and $(\sigma_8, \Sigma m_{\nu})$ planes, where the thin blue
(thick red) lines correspond to constraints in the flat (non-flat) scenario. The ``+'' (``x'') marks the mean values
of the pair in the flat (non-flat) scenario. \emph{\textbf{Right panel}}: contours in the $(\Omega_k, \Sigma m_{\nu})$
plane for the non-flat scenario. The ``x''  marks the mean values of the $(\Omega_k, \Sigma m_{\nu})$ pair.}
\label{fig:LCDM}
\end{figure*}

\begin{figure*}[t]
\centering $\begin{array}{cc}
\includegraphics[width=0.4\textwidth]{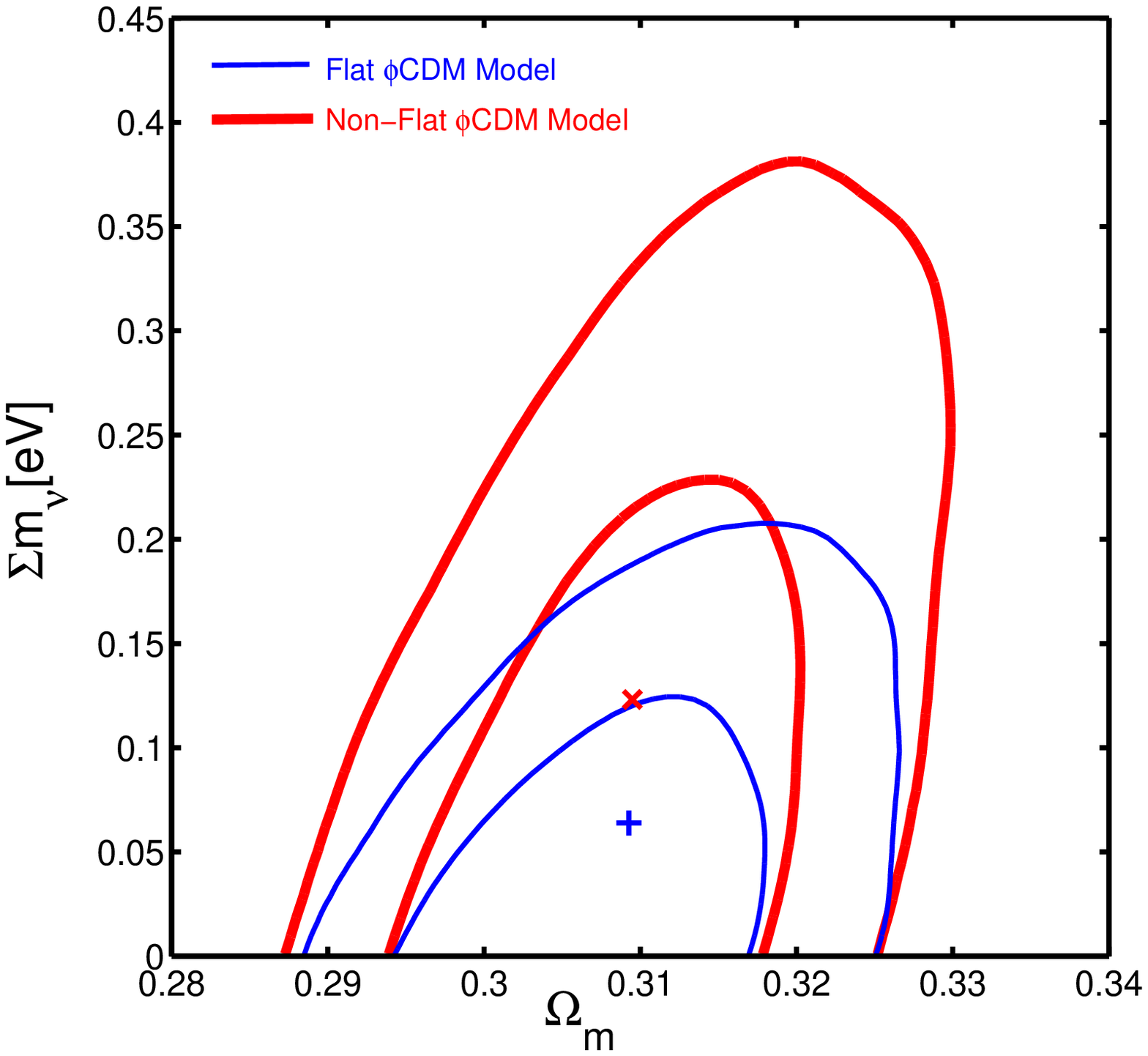},
\includegraphics[width=0.4\textwidth]{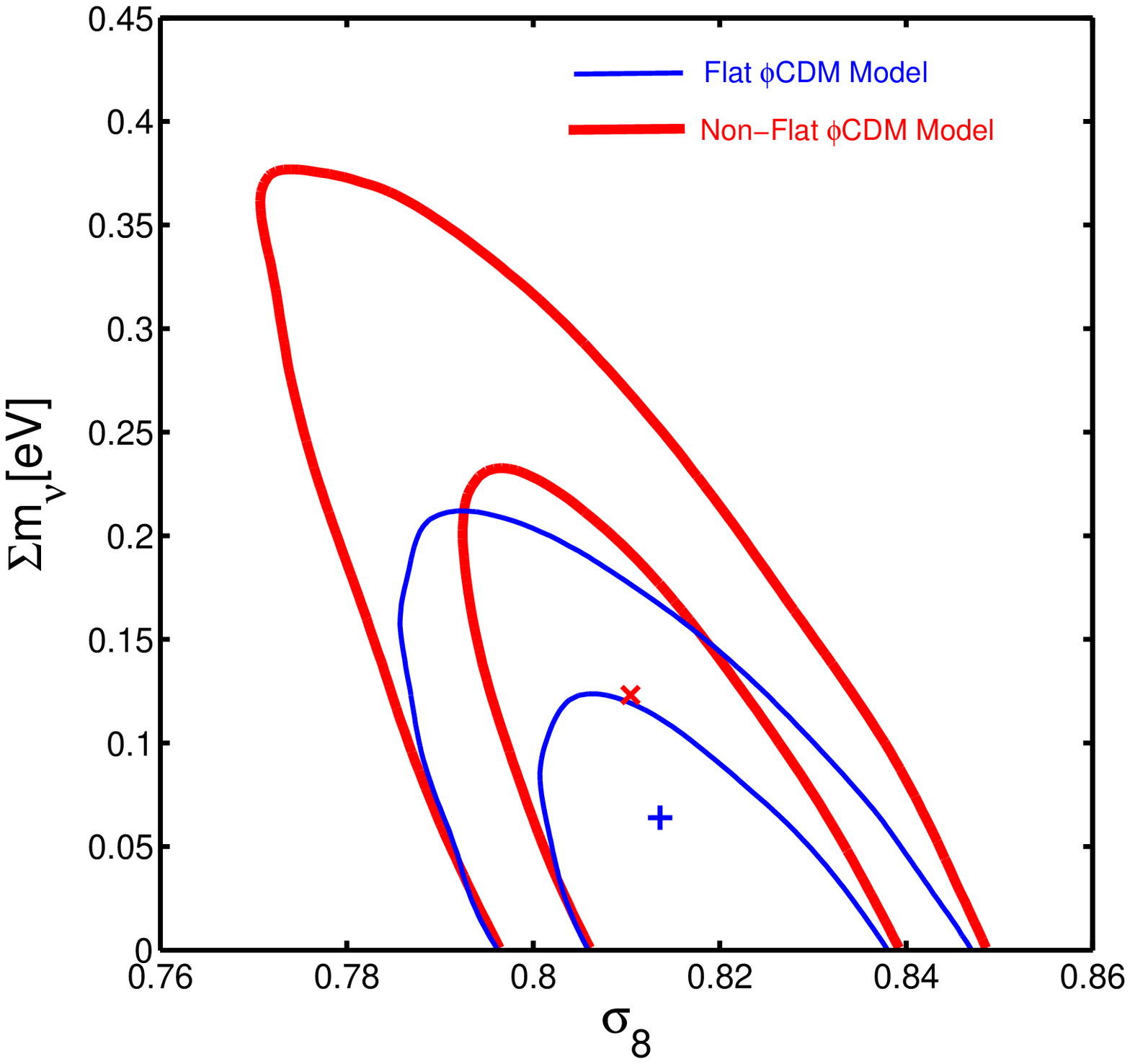}\\
\includegraphics[width=0.4\textwidth]{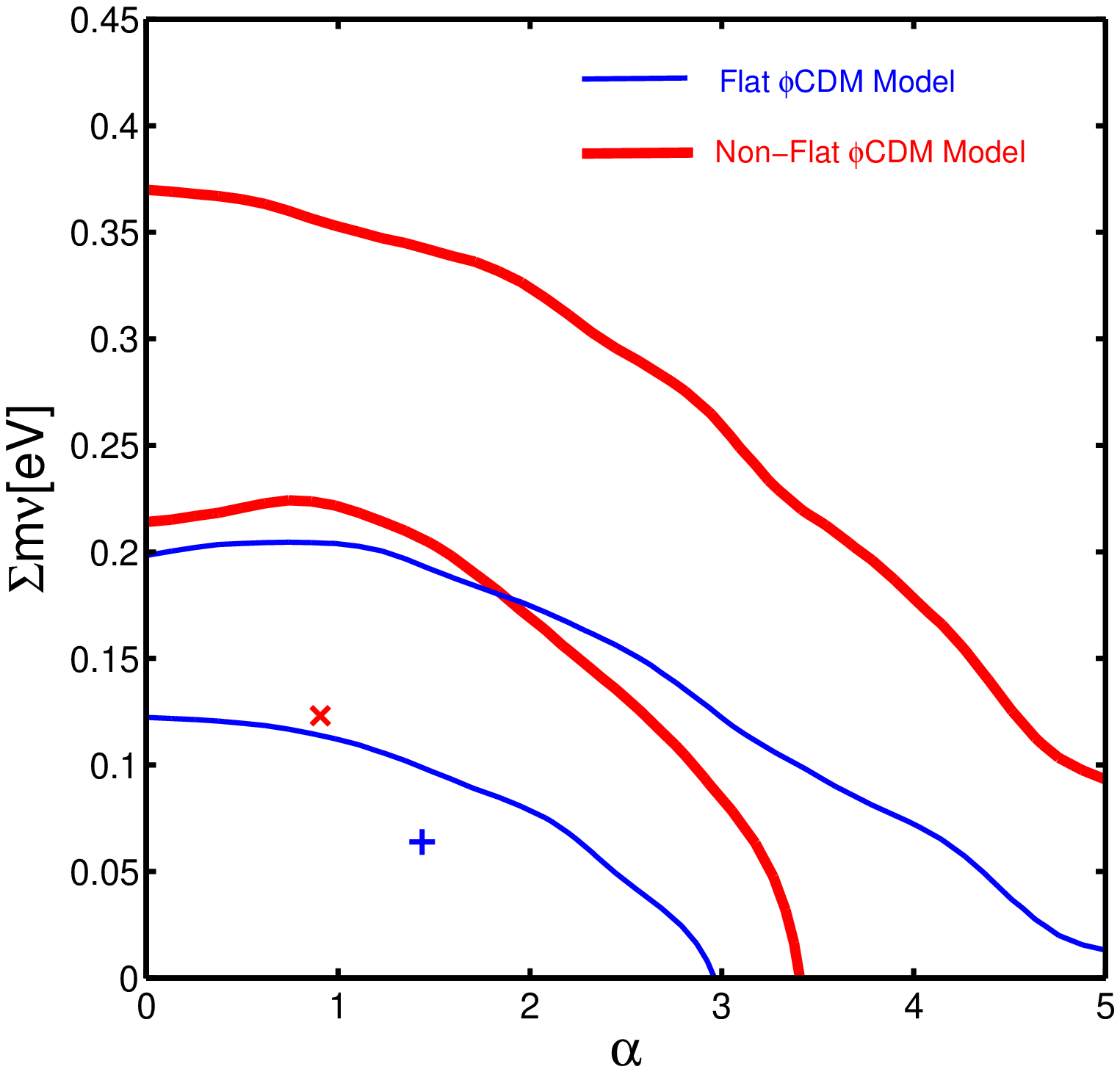},
\includegraphics[width=0.4\textwidth]{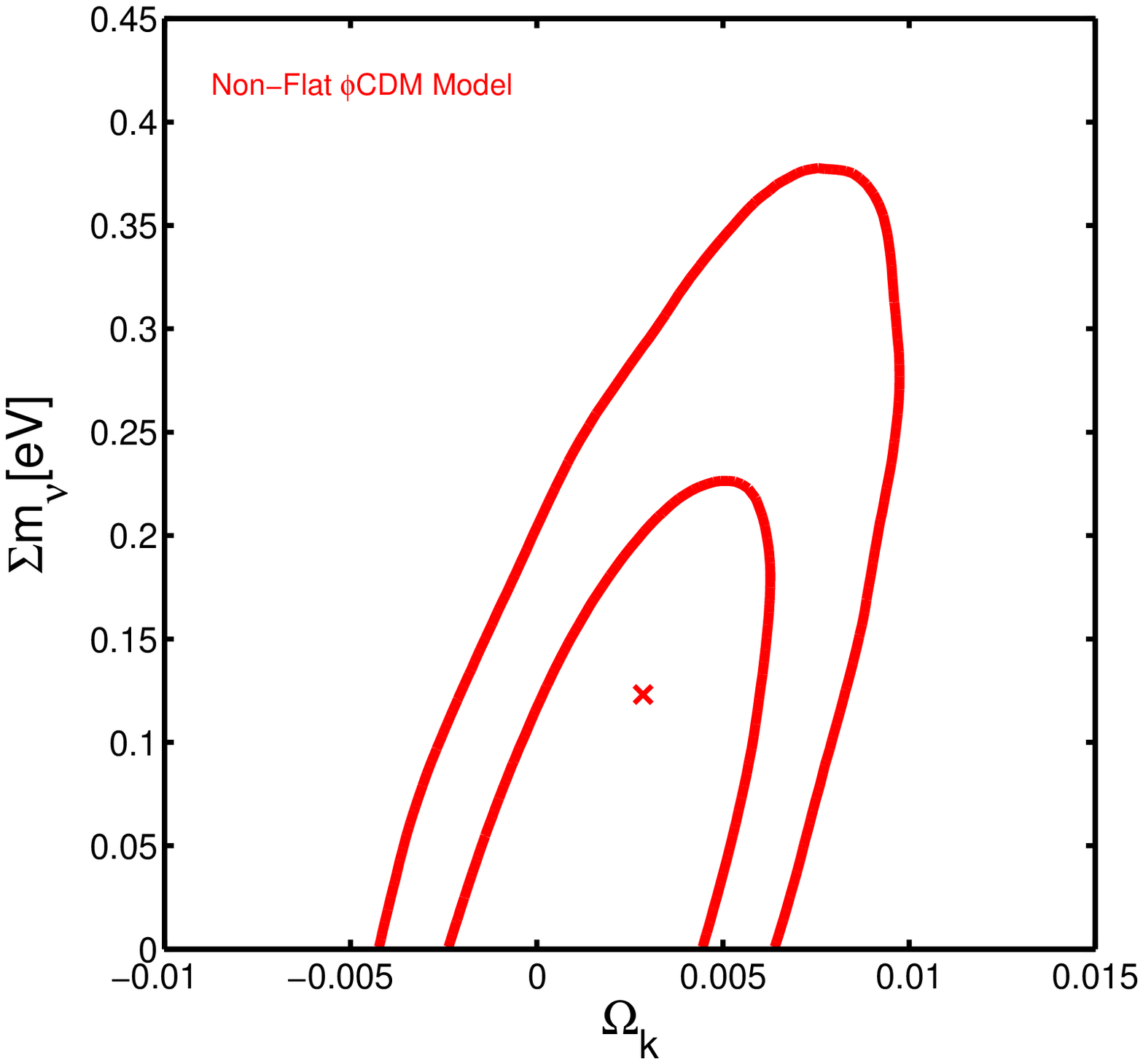}
\end{array}$
 \caption{Contours refer to the marginalized likelihoods at 68\% and 95\% confidence levels constrained from the joint analysis using the HST $H_0$ prior
 for the $\phi$CDM model  in the scenario assuming three species of degenerate massive neutrinos.
 \emph{\textbf{Upper left, upper right and lower left panels}}: contours
 in the $(\Omega_m, \Sigma m_{\nu})$, $(\sigma_8, \Sigma m_{\nu})$ and $(\alpha, \Sigma m_{\nu})$ planes, where the
 thin blue (thick red) lines correspond to constraints in the flat (non-flat) scenario. The ``+'' (``x'') marks the
 mean values of the pair in the flat (non-flat) scenario. \emph{\textbf{Lower right panel}}: contours in
 the $(\Omega_k, \Sigma m_{\nu})$ plane for the non-flat scenario. The ``x''  marks the mean values of the
 $(\Omega_k, \Sigma m_{\nu})$ pair. }
\label{fig:phiCDM}
\end{figure*}

First, we give constraints assuming three species of degenerate massive neutrinos.
Two-dimensional contours for the cosmological parameters of interest
are shown in Fig. \ref{fig:LCDM} for the flat and non-flat $\Lambda$CDM models
and in Fig.  \ref{fig:phiCDM} for the flat and non-flat $\phi$CDM models.
In these two figures HST value of $H_0$ was assumed as a prior.
One can see that constraints from the
joint data sample are quite restrictive, though there are degeneracies
between some parameters. Moreover, it turns out that with $\Omega_k$ as a free parameter the
ranges of allowed values for other parameters (except $\Omega_b h^2$ and
$100\theta_{MC}$ ) are all significantly broadened for both $\Lambda$CDM and
$\phi$CDM models.

In order to investigate the impact of the neutrino mass hierarchy, we compare the constraint results based on
 two different scenarios of the neutrino mass hierarchy as mentioned previously.
Hereafter, the scenario of assuming three species of degenerate massive neutrinos will be quoted as ``Scenario I'' for short.
And the scenario of assuming the total neutrino mass dominated by the heaviest neutrino mass
 eigenstate will be quoted as ``Scenario II''.  Corresponding mean values of the parameters
 of interest together with their 95\% confidence limits constrained from the joint analysis using the HST $H_0$ prior
are presented in Table \ref{tab:Result_lcdm} for the flat and non-flat $\Lambda$CDM models
and in Table \ref{tab:Result_phicdm} for the flat and non-flat $\phi$CDM models.
It turns out that the constraints on $\Omega_b h^2$, $\Omega_c h^2$, $100\theta_{MC}$,  $\tau$,
${\rm{ln}}(10^{10} A_s)$, $n_s$, $\Omega_m$, $\sigma_8$ and $H_0$
in the four models with different neutrino mass scenarios
are consistent with each other at 95\% CL.
In the spatially flat case, we have $\Sigma m_\nu < 0.165 (0.166)$ eV at 95\% CL in ``Scenario I'' (``Scenario II'') for the $\Lambda$CDM model,
and $\Sigma m_\nu < 0.164 (0.164)$ eV at 95\% CL in ``Scenario I'' (``Scenario II'') for the $\phi$CDM model.
In the spatially non-flat case, we have $\Sigma m_\nu < 0.299(0.354)$ eV at 95\% CL in ``Scenario I'' (``Scenario II'')
for the $\Lambda$CDM model, and $\Sigma m_\nu < 0.301 (0.364)$ eV at 95\% CL in ``Scenario I'' (``Scenario II'') for the $\phi$CDM model.
The results show that different neutrino mass scenarios just result in insignificant differences between the bounds on $\Sigma m_\nu$
for both the $\Lambda$CDM and $\phi$CDM models in the spatially flat case;
however, in the spatially non-flat case, the corresponding differences are larger than those in the spatially flat case, and
the allowed scale of $\Sigma m_\nu$ in the ``Scenario II'' is a bit larger than that in the ``Scenario I''.

Let us focus on the constraints on $\Sigma m_\nu$ and $\Omega_k$.
In ``Scenario I'', the limits at 95\% CL on the sum of neutrino masses
are $\Sigma m_\nu < 0.165 (0.299)$ eV  for the flat (non-flat) $\Lambda$CDM
model, and $\Sigma m_\nu < 0.164 (0.301)$ eV for the flat (non-flat)
$\phi$CDM model. It shows that with $\Omega_k$ as a free parameter the
95\% upper limit on $\Sigma m_{\nu}$ is about double that in the flat case
for both the $\Lambda$CDM and $\phi$CDM models. One can obtain the same conclusion in ``Scenario II''.
 The strong correlation
between $\Omega_k$ and $\Sigma m_{\nu}$ is because that the massive neutrinos
are still relativistic until recombination so they act as an additional
radiative component. And the constraint results also demonstrate that the spatially flat universe is still highly preferred.

\begin{figure*}[t]
\centering $\begin{array}{ccc}
\includegraphics[width=0.32\textwidth]{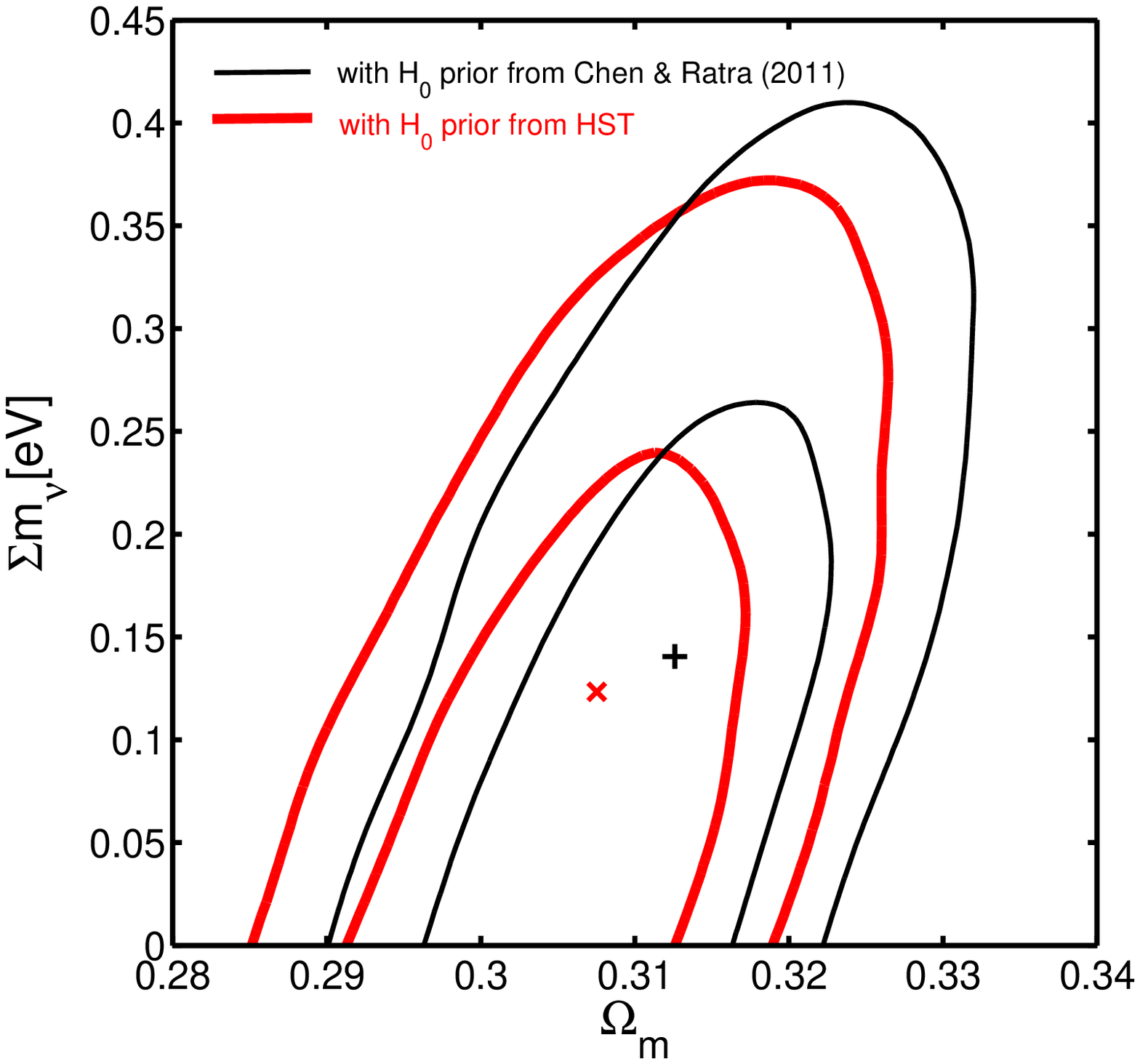}
\includegraphics[width=0.32\textwidth]{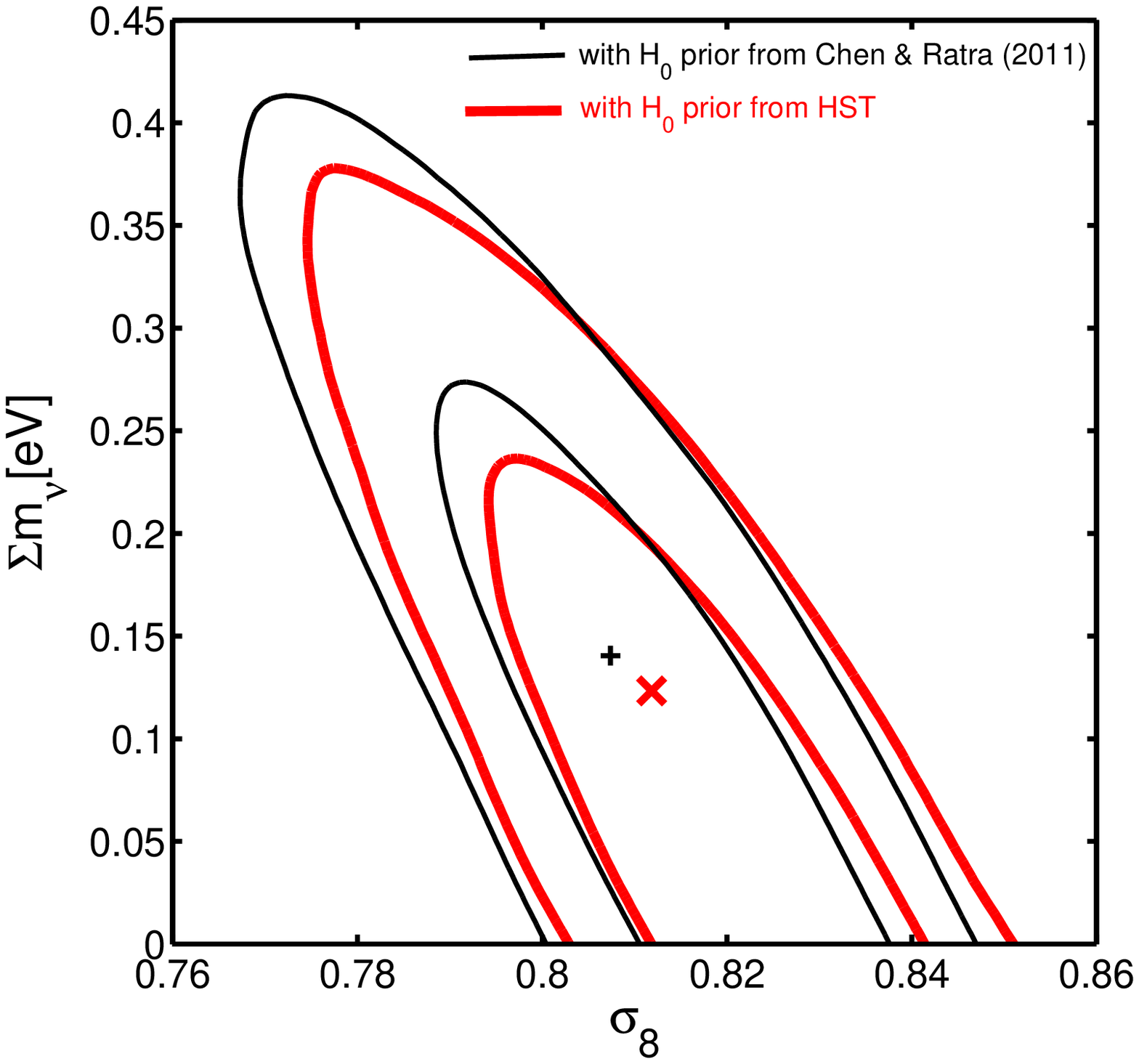}
\includegraphics[width=0.32\textwidth]{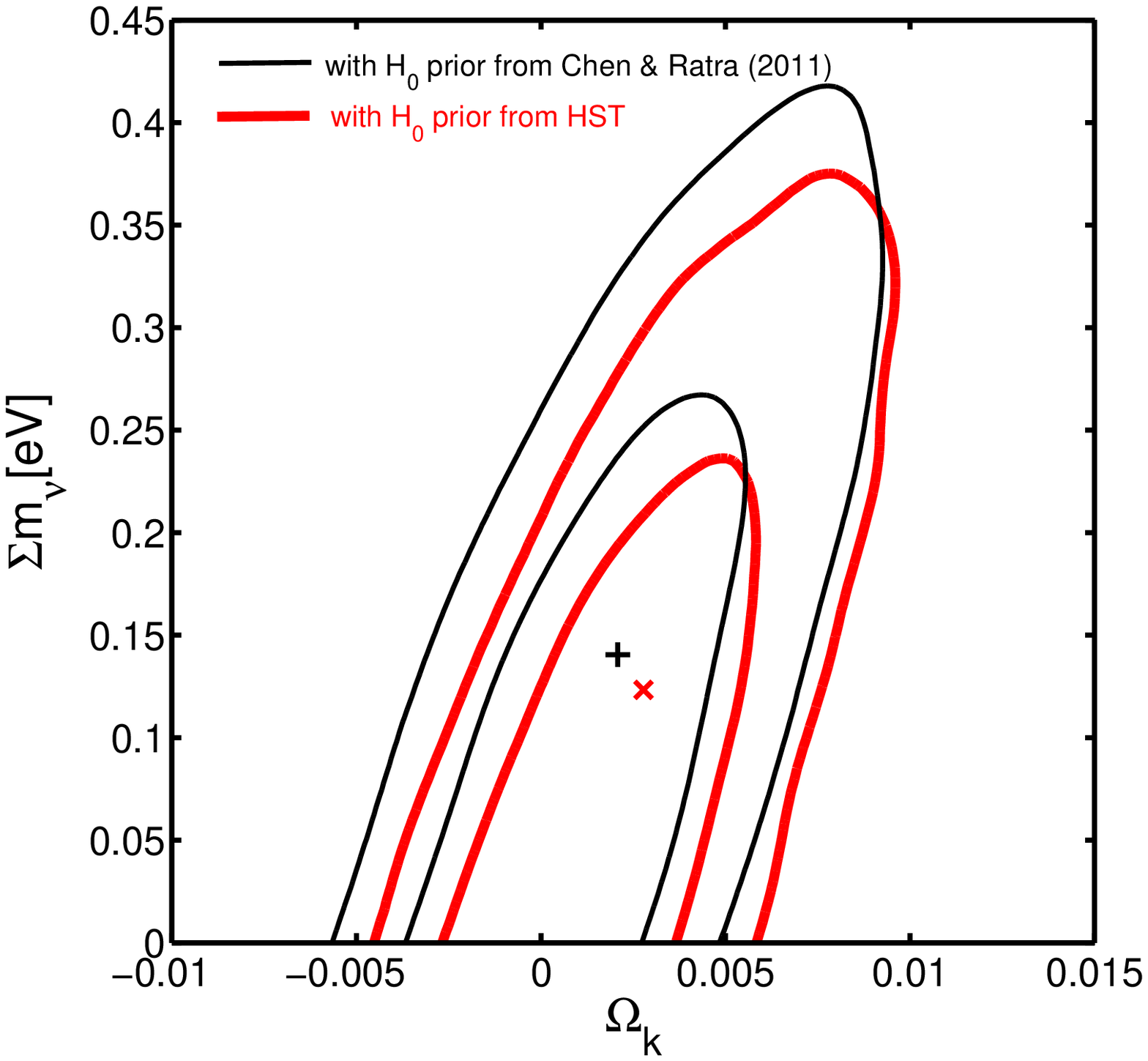}
\end{array}$
 \caption{
Contours refer to the marginalized likelihoods at 68\% and 95\% confidence levels in the non-flat $\Lambda$CDM model assuming three species of degenerate massive neutrinos
 constrained from the joint sample with two different $H_0$ priors. From left to right, contours in the $(\Omega_m, \Sigma m_{\nu})$,
  $(\sigma_8, \Sigma m_{\nu})$ and  $(\Omega_k, \Sigma m_{\nu})$ planes are presented, respectively. The thin black lines correspond
   to constraints from the joint sample with the $H_0 = (68 \pm 2.8)$ km s$^{-1}$ Mpc$^{-1}$ prior from Chen \& Ratra (2011). The thick red lines correspond to constraints from the joint sample with the $H_0 = (73.8\pm
2.4)$ km s$^{-1}$ Mpc$^{-1}$ (Riess et al. 2011) prior from HST observations. The ``+'' marks the mean values of the corresponding pair
 with $H_0$ prior from Chen \& Ratra (2011). The ``x'' marks the mean values with $H_0$ prior from Riess
et al. (2011).}
\label{fig:LCDM_mix_H0}
\end{figure*}

In order to explore the impact of the prior value of the Hubble constant
$H_0$ on the cosmological parameter estimation, we compare the constraints
resulting from the joint data sample with two different $H_0$ priors in the
non-flat $\Lambda$CDM model assuming three species of degenerate massive neutrinos. One is from HST observation with $H_0 = (73.8\pm
2.4)$ km s$^{-1}$ Mpc$^{-1}$ (Riess et al. 2011) which is used above, and
another is from the median statistics analysis of Chen \& Ratra (2011) with
$H_0 = (68 \pm 2.8)$ km s$^{-1}$ Mpc$^{-1}$. Two-dimensional confidence contours
for the cosmological parameters of interest are shown in
Fig. \ref{fig:LCDM_mix_H0} for the non-flat $\Lambda$CDM model with the two
different $H_0$ priors. One can see that the prior value of the Hubble
constant $H_0$ affects cosmological parameter estimation, but not very
significantly. In our combined analysis it is because of the weight of the other data used.
However, one can notice a certain trend, namely with smaller values of the
$H_0$ prior, the upper limit on $\Sigma m_{\nu}$ gets larger. This
implies that the parameters $H_0$ and $\Sigma m_{\nu}$ are negatively
correlated (Komatsu et al. 2009; Chen \& Xu 2016). Our result is consistent with that of Di Valentino et al. (2016) who conclude
that the bounds on the neutrino parameters may differ appreciably depending on the prior values of low redshift
quantities, such as the Hubble constant, the cluster mass bias, and the reionization optical depth.

\section{Conclusion}
\label{summary}
We have studied the $\Lambda$CDM and $\phi$CDM models with massive neutrinos assuming two different neutrino mass hierarchies
in both the spatially flat and non-flat scenarios. In the $\phi$CDM model
under consideration, the dark energy scalar field $\phi$ with an inverse
power-law potential $V(\phi)\propto \phi^{-\alpha}$ ($\alpha>0$) powers the
late-time accelerated cosmological expansion. In order to constrain model parameters, we performed a joint analysis on the data including Planck 2015 data comprising temperature and polarization of CMB anisotropies as well as CMB lensing,
 BAO data from 6dFGS, SDSS-MGS, BOSS-LOWZ and
CMASS-DR11, the JLA compilation of Type Ia supernova observations, and the
$H_0$ prior according to HST or median statistics.
The results indicate that constraints on the cosmological parameters from
this combination of data are quite restrictive. We find that the constraints
on the parameters are much tighter than those in the previous literature
(Chen \& Xu 2016), which made use of a combination of the CMB temperature
power spectrum likelihoods from Planck 2013 and the CMB polarization power
spectrum likelihoods from nine-year WMAP (WMAP9), the galaxy clustering data
from WiggleZ and BOSS DR11, and the JLA compilation of Type Ia supernova
observations. More recent paper by Chen \& Xu (2016) studying
the $\Lambda$CDM and $\phi$CDM
models with massive neutrinos assumed only the spatially flat case.

The results of our paper clearly show that cosmological bounds
on the total neutrino mass $\Sigma m_{\nu}$ are very tight, however,
they are significantly correlated with the curvature term. It turns out that
with $\Omega_k$ as a free parameter the 95\% upper limit on $\Sigma m_{\nu}$
is relaxed by more than a factor of two with respect to that in the flat case
for both the $\Lambda$CDM and $\phi$CDM scenarios. Furthermore, the bounds on $\Sigma m_{\nu}$ based on two
 different neutrino mass hierarchies have insignificant differences
in the spatially flat case for both the $\Lambda$CDM and $\phi$CDM models, however, the corresponding differences are
larger in the non-flat case.
Moreover, for a given neutrino mass hierarchy, the bounds
on $\Sigma m_{\nu}$ in $\Lambda$CDM and $\phi$CDM scenarios have small
differences, irrespective of whether
$\Omega_k$ is fixed at zero or is it taken as a free parameter.
For example, in the scenario of assuming three species of degenerate massive neutrinos, when $\Omega_k = 0$,
 we have $\Sigma m_\nu < 0.165 (0.164)$ eV
at 95\% CL for the $\Lambda$CDM ($\phi$CDM) model; when $\Omega_k \neq 0$, we
have $\Sigma m_\nu < 0.299(0.301)$ eV  at 95\% CL for the $\Lambda$CDM
($\phi$CDM) model.
Additionally, in the scenario assuming three species of degenerate massive neutrinos, we find $\alpha < 3.494\;(3.938)$ at 95\% CL for
the flat (non-flat) $\phi$CDM model, while the $\Lambda$CDM scenario corresponding to $\alpha = 0$ is not ruled out at this confidence level.
One can obtain the same conclusion in the scenario assuming the total neutrino mass dominated by the heaviest neutrino mass eigenstate.
In general, the constraints on the cosmological parameters are similar in the $\Lambda$CDM and $\phi$CDM
 models, and the bounds on the total neutrino mass $\Sigma m_{\nu}$ are not that sensitive to
the underlying cosmological models under consideration. Massive neutrinos mainly affect the redshift of matter-radiation equality $z_{eq}$ (and also being
relativistic at the $z_{eq}$ they are counted as non-relativistic now thus being entangled with $\Omega_c h^2$). At this epoch neither $\Lambda$ nor $\phi$ contribute significantly to the background expansion. 
 Consequently,these results imply that the
observational data that we have employed here still cannot distinguish
whether dark energy is a time-independent cosmological constant or varies
mildly in space and slowly in time.

%%%%%%%%%%%%%%%%%%%%%%%%%%%%%%%%%%%%%%%%%%%%%%%%%%%%%%%%%%%%%%%%%%%
%%%%%%%%%%%%%%%%%%%%%%%%%%%%%%%%%%%%%%%%%%%%%%%%%%%%%%%%%%%%%%%%%%%
%%%%%%%%%%%%%%%%%%%%%%%%%%%%%%%%%%%%%%%%%%%%%%%%%%%%%%%%%%%%%%%%%%%
\acknowledgments
Yun Chen would like to thank Jun-Qing Xia for useful discussions. Y.C. was supported by the National Natural Science Foundation of China (Nos. 11133003 and 11573031), the China Postdoctoral Science Foundation (No. 2015M571126),
and the Young Researcher Grant of National Astronomical Observatories, Chinese Academy of Sciences. B.R. was supported in part by DOE grant DEFG 03-99EP41093. M.B. was supported by the Polish NCN grant
DEC-2013/08/M/ST9/00664 and the Poland-China Scientific and Technological Cooperation Committee
Project No. 35-4. M.B. obtained approval of foreign talent introducing project in China and
gained special fund support of foreign knowledge introducing project.
S.L. was supported by the National Natural Science Foundation of China under Grants No. 11347163, and
the Science and Technology Program Foundation of the Beijing Municipal Commission of Education of China under Grant No. KM201410028003.
Z.H.Z. was supported by the Chinese Ministry of Science and Technology National Basic Science Program
(Project 973) under Grant No.2012CB821804 and
2014CB845806. Y.C. and Z.H.Z. were also supported by the Strategic Priority Research Program
``The Emergence of Cosmological Structure'' of the Chinese
Academy of Sciences (No. XDB09000000).

%%%%%%%%%%%%%%%%%%%%
%figure1
%%%%%%%%%%%%%%%%%%%%

%%%%%%%%%%%%%%%%%%%%%%%%
%figure2
%%%%%%%%%%%%%%%%%%%%%

\clearpage
%%%%%%%%%%%
%%%Table 1
%%%%%%%%%%%
\begin{table*}
%\vspace*{20mm}
%\hspace*{10mm}
\begin{center}{\large%\tiny %\scriptsize%\footnotesize%\small %\large
%\begin{tabular}{|c|c|c|c|c|}
\begin{tabular}{|p{3cm}|p{3cm}|p{3cm}|p{3cm}|p{3cm}|}\hline

& \multicolumn{4}{c|} {$\Lambda$CDM model} \\
 \cline{2-5}
Parameters & \multicolumn{2}{c|}{Scenario I} & \multicolumn{2}{c|}{Scenario II} \\
 \cline{2-5}
 % \hline
 & Flat& Non-Flat & Flat & Non-Flat\\

\hline
$\Omega_b h^2$ &  $0.0223 \pm 0.0003$ & $0.0222 \pm 0.0003$ & $0.0223 \pm 0.0003$ & $ 0.0222 \pm 0.0003$ \\
$\Omega_c h^2$ & $0.1184 \pm 0.0021$  & $0.1195^{+0.0030}_{-0.0029}$  & $0.1184 \pm 0.0021$ &  $0.1196 \pm 0.0030$ \\
$100\theta_{MC}$ & $1.0410 \pm 0.0006$ & $1.0408 \pm 0.0006$ & $1.0410 \pm 0.0006$ &  $1.0408 \pm 0.0006$ \\
$\tau$          & $0.0676^{+0.0289}_{-0.0260}$ &  $0.0715^{+0.0326}_{-0.0287}$& $0.0685^{+0.0279}_{-0.0260}$ &  $0.0739^{+0.0322}_{-0.0309}$  \\
${\rm{ln}}(10^{10} A_s)$ &  $3.0664^{+0.0537}_{-0.0488}$ &  $3.0767^{+0.0643}_{-0.0557}$ & $3.0679^{+0.0520}_{-0.0486}$ & $3.0812^{+0.0621}_{-0.0594}$  \\
$n_s$ &  $0.9675^{+0.0082}_{-0.0080}$ &  $0.9650\pm 0.0095$ & $0.9674^{+0.0079}_{-0.0078}$ &  $0.9642^{+0.0097}_{-0.0100}$  \\
$\Omega_k$ &   ... &  $0.0028^{+0.0055}_{-0.0051}$ & ... &  $0.0033^{+0.0058}_{-0.0051}$ \\
$\Sigma m_\nu$ (eV) &   $<0.165$ &  $< 0.299$ & $< 0.166$ &  $< 0.354$  \\
\hline
$\Omega_m$ &  $0.307^{+0.014}_{-0.013}$ &  $0.308^{+0.016}_{-0.015}$ &  $0.308^{+0.014}_{-0.013}$ &  $0.309 \pm 0.016 $  \\
$\sigma_8$  & $0.816^{+0.022}_{-0.024}$ &  $0.812 ^{+0.028}_{-0.030}$ & $0.815^{+0.023}_{-0.024}$ &  $0.807^{+0.032}_{-0.037}$ \\
$H_0$ (km/s/Mpc)&  $67.87^{+1.05}_{-1.11}$ &  $68.22^{+1.43}_{-1.38}$ &  $67.83^{+1.03}_{-1.12}$ &  $ 68.18^{+1.36}_{-1.38}$  \\

\hline

\end{tabular}}\\
\end{center}
\caption{Constraints from the joint analysis using the HST $H_0$ prior, for the $\Lambda$CDM model in spatially flat and non-flat cases
with two different scenarios for the neutrino mass hierarchy. The ``Scenario I'' and ``Scenario II'' denote two different scenarios of
the neutrino mass hierarchy, the implications of which are described in Sec. {\ref{Results}}.
We present the mean values with 95\% confidence limits for the parameters of interest.
 The top block contains parameters with uniform priors that are varied in the MCMC chains. The lower block shows various
 derived parameters.}\label{tab:Result_lcdm}
\end{table*}

%%%%%%%%%%%
%%%Table 2
%%%%%%%%%%%
\begin{table*}
%\vspace*{20mm}
%\hspace*{10mm}
\begin{center}{\large%\tiny %\scriptsize%\footnotesize%\small %\large
%\begin{tabular}{|c|c|c|c|c|}
\begin{tabular}{|p{3cm}|p{3cm}|p{3cm}|p{3cm}|p{3cm}|}\hline

& \multicolumn{4}{c|} {$\phi$CDM model} \\
 \cline{2-5}
Parameters & \multicolumn{2}{c|}{Scenario I} & \multicolumn{2}{c|}{Scenario II} \\
 \cline{2-5}
 % \hline
 & Flat& Non-Flat & Flat & Non-Flat\\

\hline
$\Omega_b h^2$  &  $0.0223 \pm 0.0003$ & $0.0222 \pm 0.0003$ &  $0.0223 \pm 0.0003$ & $0.0222 \pm 0.0003$\\
$\Omega_c h^2$ &  $0.1183 \pm 0.0021$ &  $0.1196 \pm 0.0030$ &  $0.1183^{+0.0021}_{-0.0022}$ &  $0.1196 \pm 0.0030$ \\
$100\theta_{MC}$ & $1.0410 \pm 0.0006$ & $1.0408 \pm 0.0007$ & $1.0410 \pm 0.0006$ & $1.0408 \pm 0.0007$ \\
$\tau$          & $0.0685^{+0.0283}_{-0.0263}$ &  $0.0722^{+0.0330}_{-0.0313}$  & $0.0699^{+0.0283}_{-0.0262}$ &  $0.0748^{+0.0319}_{-0.0298}$\\
${\rm{ln}}(10^{10} A_s)$ &$3.0679^{+0.0533}_{-0.0492}$ & $3.0782^{+0.0642}_{-0.0601}$  &  $3.0703^{+0.0526}_{-0.0488}$ &  $3.0831^{+0.0616}_{-0.0567}$ \\
$n_s$ & $0.9680^{+0.0081}_{-0.0080}$ &  $0.9647^{+0.0096}_{-0.0092}$ &  $0.9678^{+0.0083}_{-0.0081}$ &  $0.9643\pm 0.0097$  \\
$\Omega_k$ & ... &  $0.0031^{+0.0056}_{-0.0049}$ &   ... &  $0.0036^{+0.0059}_{-0.0055}$ \\
$\Sigma m_\nu$ (eV) & $< 0.164$ &  $< 0.301$ &   $<0.164$ &  $< 0.364$ \\
$\alpha$ & $<3.494$ & $ < 3.938$ & $<3.425$ &  $ < 3.941$  \\
\hline
$\Omega_m$ &  $0.309\pm 0.015$ &  $0.311^{+0.017}_{-0.015}$ &  $0.310^{+0.015}_{-0.014}$ &  $0.311 \pm 0.017$ \\
$\sigma_8$ & $0.814^{+0.023}_{-0.024}$ &  $0.809^{+0.028}_{-0.031}$ & $0.813^{+0.023}_{-0.025}$ &  $0.805 ^{+0.033}_{-0.038}$  \\
$H_0$ (km/s/Mpc) &  $67.61^{+1.24}_{-1.34}$ &  $67.89^{+ 1.49}_{-1.50}$ &  $67.57^{+1.20}_{-1.33}$ &  $ 67.91^{+1.45}_{-1.50}$ \\

\hline

\end{tabular}}\\
\end{center}
\caption{Constraints from the joint analysis using the HST $H_0$ prior, for the $\phi$CDM model in spatially flat and non-flat cases
with two different scenarios for the neutrino mass hierarchy. The mean values with 95\% confidence limits for the parameters of interest are displayed. The top block contains parameters with uniform priors that are varied in the MCMC chains. The lower block shows various
derived parameters. The implications of ``Scenario I'' and ``Scenario II''are the same as those in Table \ref{tab:Result_lcdm}.}
\label{tab:Result_phicdm}
\end{table*}

%%%%%%%%%%%%%%%%%%%%%%%%
%figure3
%%%%%%%%%%%%%%%%%%%%%


\begin{thebibliography}{}

\bibitem{Abe2008}Abe, S., et al., KamLAND Collaboration, Phys. Rev. Lett. 100 (2008) 221803 [arXiv:0801.4589]
\bibitem{Adam2015} Adam, R., et al., Planck Collaboration, ArXiv e-prints, arXiv:1502.01582
\bibitem{Ade2015}Ade, P.~A.~R., et al., Planck Collaboration, ArXiv e-prints, arXiv:1502.01589
\bibitem{Agafonova2010}Agafonova, N., et al., OPERA Collaboration, 2010, Phys. Lett. B, 691, 138 [arXiv:1006.1623]
\bibitem{Ahmad2001}Ahmad, Q. R., et al., SNO Collaboration, 2001, Phys. Rev. Lett., 87, 071301 [arXiv:nucl-ex/0106015]
\bibitem{Ahn2012}Ahn, J. K., et al., RENO collaboration, 2012, Phys. Rev. Lett., 108, 191802 [arXiv:1204.0626]
\bibitem{An2012}An, F. P., et al., Daya Bay Collaboration, 2012,  Phys. Rev. Lett., 108, 171803 [arXiv:1203.1669]
\bibitem{Anderson2014}Anderson, L., Aubourg, {\'E}., Bailey, S., et al., 2014, MNRAS, 441, 24 [arXiv:1312.4877]
\bibitem{Ashie2005}Ashie, Y., et al., Super-Kamiokande Collaboration, Phys. Rev. D 71 (2005) 112005 [arXiv:hep-ex/0501064]
\bibitem{Aubourg2015}Aubourg, E., et al. 2015, Phys. Rev. D, 92, 123516 [arXiv:1411.1074]
\bibitem{Avsajanishvili2014}Avsajanishvili, O., Arkhipova, N.~A., Samushia, L.,
         \&  Kahniashvili, T., 2014, Eur. Phys. J. C 74, 3127 [arXiv:1406.0407]
\bibitem{Avsajanishvili2015}Avsajanishvili, O., Samushia, L., Arkhipova, N.~A.,
         \& Kahniashvili, T., 2015, arXiv:1511.09317
\bibitem{Balantekin2013} A. B. Balantekin,\& W.C. Haxton, Prog. Part. Nucl. Phys. 71 (2013) 150 [arXiv:1303.2272]
\bibitem{Bennett2003} Bennett, C. L., et al., Astrophys. J. Suppl. 148 (2003) 1
\bibitem{Betoule2014}Betoule, M., et al., 2014, Astron. Astrophys. 568, A22 [arXiv:1401.4064]
\bibitem{Beutlerf2011}Beutler, F., Blake, C., Colless, M., et al.,  2011, MNRAS, 416, 3017 [arXiv:1106.3366]
\bibitem{Calabrese2012}Calabrese, E., Archidiacono, M., Melchiorri, A., \&
         Ratra, B. 2012, Phys. Rev. D, 86, 043520 [arXiv:1205.6753].
\bibitem{Chae2004}Chae, K.-H., Chen, G., Ratra, B., Lee, D.-W. 2004,
         Astrophys. J., 607, L71 [arXiv:astro-ph/0403256]
\bibitem{Chen2004}Chen, G. \& Ratra, B. 2004,  Astrophys. J., 612, L1
         [arXiv:astro-ph/0405636]
\bibitem{Chen2011}Chen, G. \& Ratra, B. 2011, PASP, 123, 1127 [arXiv:1105.5206]
\bibitem{Chen2013}Chen, Y. \& Ratra, B. 2013,  Phys. Lett. B, 703, 406
         [arXiv:1106.4294]
\bibitem{chen2015}Chen, Y., et al., 2015, J. Cosmol. Astropart. Phys. 02, 010 [arXiv:1312.1443]
\bibitem{Chen2016}Chen, Y. \& Xu, L., 2016, Phys. Lett. B, 752, 66 [arXiv:1507.02008]
\bibitem{Di_Valentino2016} Di Valentino, E., et al., 2016, \prd, 93, 083527 [arXiv:1511.00975]
\bibitem{Farooq2013a}Farooq, O., Crandall, S., \& Ratra, B. 2013a, Phys. Lett.
         B, 726, 72 [arXiv:1305.1957]
\bibitem{Farooq2013b}Farooq, O., Mania, D., \& Ratra, B. 2013b, ApJ, 764, 138
        [arXiv:1211.4253]
\bibitem{Farooq2015}Farooq, O., Mania, D., \& Ratra, B. 2015,
         ApSS, 357, 11 [arXiv:1308.0834]
\bibitem{Farooq2013}Farooq, O. \& Ratra, B. 2013,  ApJ, 766, L7
        [arXiv:1301.5243]
\bibitem{Fogli2006} G. L. Fogli, E. Lisi, A. Marrone, \& A. Palazzo, Prog. Part. Nucl. Phys. 57 (2006) 742 [arXiv:hep-ph/0506083]
\bibitem{Freedman2001}Freedman, W. L., Madore, B. F., Gibson, B. K., et al. 2001, \apj  553, 47 [arXiv:astro-ph/0012376]
\bibitem{Fukuda1998}Fukuda, Y., et al., Super-Kamiokande Collaboration, 1998, Phys. Rev. Lett., 81, 1562 [arXiv:hep-ex/9807003]
\bibitem{Gosenca2015}Gosenca, M. \& Coles, P. 2015, arXiv:1502.04020
\bibitem{Hannestad2005}Hannestad, S., 2005, Phys. Rev. Lett. 95, 221301 [astro-ph/0505551]
\bibitem{Hu1996}Hu, W., Sugiyama, N., 1996, Astrophys. J. 471, 542 [arXiv:astro-ph/9510117]
\bibitem{Komatsu2009} Komatsu, E., et al., 2009, \apjs 180,  330 [arXiv:0803.0547]
\bibitem{Lesgourgues2006}Lesgourgues, J., Pastor, S., Phys. Rep. 429 (2006) 307¨C379 [arXiv:astro-ph/0603494]
\bibitem{Lesgourgues2012}Lesgourgues, J., Pastor, S., Adv. High Energy Phys. 2012 (2012) 608515 [arXiv:1212.6154]
\bibitem{Lewis2002}Lewis, A., Bridle, S., 2002, Phys. Rev. D 66, 103511 [arXiv:astro-ph/0205436]
\bibitem{Li2012}Li, H. \& Xia, J.-Q., 2012, \jcap  11, 039 [arXiv:1210.2037]
\bibitem{Lima2015}Lima, N.~A., Liddle, A.~R., Sahl\'en, M., \& Parkinson, D.
         2015, arXiv:1501.02678
\bibitem{Maltoni2004} M. Maltoni, T. Schwetz, M.A. Tortola, \& J.W.F. Valle, New J. Phys. 6 (2004) 122 [arXiv:hep-ph/0405172]
\bibitem{Pavlov2014}Pavlov, A., Farooq, O., Ratra, B., 2014,  Phys. Rev. D 90, 023006 [arXiv:1312.5285]
\bibitem{Pavlov2013}Pavlov, A., Westmoreland, S., Saaidi, K., Ratra, B., 2013, \prd  88, 123513 [arXiv:1307.7399]
\bibitem{Peebles1988}Peebles, P.~J.~E., Ratra, B., 1988,  Astrophys. J. 325, L17
\bibitem{Podariu2000}Podariu, S., Ratra, B., 2000 Astrophys. J. 532, 109 [arXiv:astro-ph/9910527]
\bibitem{Ratra1988}  Ratra, B., Peebles, P.~J.~E., 1988, Phys. Rev. D 37, 3406.
\bibitem{Riess2004}Riess, A. G., et al. 2004, \apj, 607, 665 [arXiv:astro-ph/0402512]
\bibitem{Riess2011}Riess, A. G., Macri, L., Casertano, S., et al., 2011, \apj 730, 119 [arXiv:1103.2976]
\bibitem{Ross2014}Ross, A. J., Samushia, L., Howlett, C., et al., 2014, ArXiv e-prints, arXiv:1409.3242
\bibitem{Samushia2007}Samushia, L., Chen, G., \& Ratra, B. 2007, arXiv:0706.1963
\bibitem{Samushia2010}Samushia, L., Ratra,  B., 2010 Astrophys. J. 714, 1347 [arXiv:0905.3836]
\bibitem{Sievers2013} Sievers, J.~L., et al. 2013, JCAP, 1310, 060 [arXiv:1301.0824]
\bibitem{Smith2012} Smith, A., et al. 2012, \prd  85, 123521 [arXiv:1112.3006]
\bibitem{Tegmark2004}Tegmark, M., et al. 2004, Phys. Rev. D 69, 103501
\bibitem{Wang2012}Wang, X., et al., 2012, \jcap 11, 018 [arXiv:1210.2136]
\bibitem{Xia2007}Xia, J.-Q., Zhao, G.-B., Zhang, X., 2007, \prd 75, 103505 [arXiv:astro-ph/0609463]
\bibitem{Xia2008}Xia, J.-Q., Li, H., Zhao, G.-B., Zhang, X., 2008, \prd  78, 083524 [arXiv:0807.3878]

\end{thebibliography}
\end{document}